# Quantum bit string commitment protocol using polarization of mesoscopic coherent states

Fábio Alencar Mendonça      *Rubens Viana Ramos

*Department of Teleinformatic Engineering, Federal University of Ceara, Campus do Pici, 725, C.P. 6007, 60755-640, Fortaleza-Ceará, Brazil*

**Abstract**

In this work, we propose a quantum bit string commitment protocol using polarization of mesoscopic coherent states. The protocol is described and its security against brute force and quantum cloning machine attack is analyzed.

Bit commitment is one of the most important cryptographic protocols that can be used to realize, among others, coin tossing, zero-knowledge proofs and electronic voting. Due to its importance, it was one of the firsts cryptographic protocols that researchers tried to find a quantum version [1-4], believing that a quantum bit commitment (QBC) protocol could be unconditionally secure. However, it was proved that it is impossible to construct an unconditionally secure QBC protocol using qubits encoded in single-photon pulses [5,6], that is, when Alice sends a single-photon to Bob. Nevertheless, this does not mean that QBC has security similar to classical versions. In fact, limiting the power of the participants, it is possible to construct QBC protocols resistant to some types of attacks [7,8].

On the other hand, aiming to improve the performance of practical quantum communication systems (limited by the use of single-photons that are hard to produce, detect and survive to the channel losses), the use of polarization of mesoscopic coherent states (MCS) as quantum information has been proposed [9-15]. In fact, the proposed quantum key expansion protocols using MCS can provide high transmission rates over long distances; however, its security has been a point of intense discussions [16-19].

______________________
*Corresponding author
*E-mail addresses*: rubens@deti.ufc.br, alencar@deti.ufc.br.

In this direction, this work proposes a quantum bit string commitment protocol using polarization of mesoscopic coherent states (QBSC_MCS). As will be seen, the proposed protocol has all the advantages of using coherent states instead of single-photon pulses and it has not the security problems of the quantum key expansion protocols proposed in [9-15] since it is not necessary that Alice and Bob share an initial key in advance.

The $n$-bit string commitment protocol can be shortly described in the following way: Alice has to choose one between (at maximum) $2^n$ bit strings. She has to convince Bob that she did her choice, at the same time she is not allowed to change her mind and Bob is not allowed to know Alice's choice without her permission. The proposed protocol can be understood observing Fig. 1.

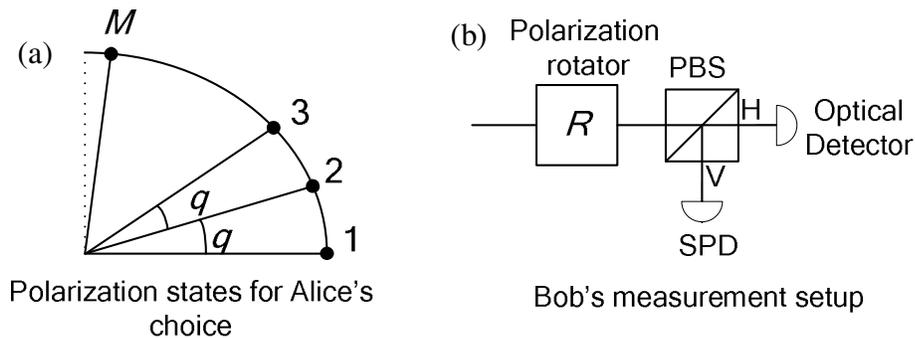

Fig. 1 – Quantum bit string commitment protocol: (a) distribution of polarization states and (b) polarization measurement setup. SPD – single-photon detector, PBS – Polarization beam splitter.

As can be seen in Fig. 1, there are $M$ polarization states and, hence, the bit string has $\log_2(M)$ bits. Supposing honest Alice and Bob, the QBSC_MCS can be described as follows: 1) Alice chooses one of the $M$ polarization states and sends it to Bob using coherent states. 2) Bob, which is supposed to have a quantum memory, waits Alice to reveal her choice. Having this information, Bob performs a measurement of the polarization using the appropriate basis in order to confirm Alice's choice. Now, supposing real Alice and Bob, some cares must be taken into account in order to avoid one of them cheating the other.

In order to do not be cheated by Alice, Bob limits the minimal mean photon number that Alice can use. For this, in his measurement setup, shown in Fig. 1(b), Bob uses a non-single-photon detector at the output where the photons are expected to emerge (once Alice revealed the polarization, Bob sets the polarization rotator in order to transform the state sent by Alice in a horizontal polarization state) and he uses a single-photon detector at the other output. In the ideal case, if detection occurs in the single-photon detector, Bob will know that Alice lied about her choice. Once Bob will confirm Alice's choice measuring the polarization with non-single-photon detector, Alice will be forced to use multi-photon pulses. On one hand, Bob limits the minimal mean photon number of the coherent state used in order to avoid Alice to choose one polarization and, at the end, to change her choice to one of the neighbors. On the other hand, Alice will use the minimal mean photon number permitted, otherwise, if Bob has enough photons, he can use the brute-force attack to determine Alice's choice without her permission.

Those non-cheating statements can be mathematically described. The inner product between two neighbors polarization is given by:

$$|\langle \alpha, \beta | R(\theta) | \alpha, \beta \rangle|^2 = \exp\left[-4(\alpha^2 + \beta^2)\sin^2\left(\frac{\theta}{2}\right)\right] \quad (1)$$

$$R(\theta) = \begin{bmatrix} \cos(\theta) & -\sin(\theta) \\ \sin(\theta) & \cos(\theta) \end{bmatrix}. \quad (2)$$

In (1), without loss of generality, we have used the two-mode coherent state $|\alpha,\beta\rangle$ with $\alpha$ and $\beta$ real numbers. Thus, we are considering only linear polarizations. In order to analyze the security, initially we consider the parameter, $r_s^1$ given by:

$$r_s^1 = \exp\left[-4\langle n \rangle \sin^2\left(\frac{\pi}{4M}\right)\right]. \quad (3)$$

In (3) we have used $\langle n \rangle = \alpha^2 + \beta^2$ and $\theta = \pi/(2M)$. Thus, $r_s^1$ shows the amount that two neighbors polarizations fails to be orthogonal. Since $M$ is a given parameter of the protocol, Alice and Bob agree in a value for $r_s^1$ and, using (3), they find the mean photon number to be used. Bob sets his optical receiver to have a good signal-noise relation with optical pulses having mean photon number equal to $\langle n \rangle$ because this is the minimal mean photon number that Alice must use.

In order to check the security against some attacks, let us consider the brute-force attack and the quantum cloning machine attack. In the first case, Bob splits the pulse sent by Alice in $M$ pulses having mean photon number equal to $\langle n \rangle/M$, and he tests each pulse in a different polarization basis, now using single-photon detectors in both PBS's output.

The probability of Bob to identify the polarization sent by Alice, without any doubt (detection in both detectors for all basis different of the correct one), is given by:

$$P_b = \left(1 - e^{-\frac{\langle n \rangle}{M}}\right) \sum_{k=1}^{M} P_A(\theta_k) \prod_{\substack{i=1 \\ (i \neq k)}}^{M} \left\{ 1 - 2e^{-\frac{\langle n \rangle}{2M}} \cosh\left[\frac{\langle n \rangle}{2M} \cos\left[2(\theta_k - \theta_i)\right]\right] + e^{-\frac{\langle n \rangle}{M}} \right\}. \quad (4)$$

In (4) $\theta_k$ is the polarization state chosen by Alice and $P_A(\theta_k)$ is the probability of $\theta_k$ to be chosen by Alice.

On the other hand, in the quantum cloning machine attack, Bob splits the pulse sent by Alice in $N$ ($1 \leq N \leq M-1$) pulses and he uses a quantum cloning machine of coherent states (QCM_CS) [20-26] to produce $M$ copies. In order to make the cloning, Bob firstly separates the horizontal and vertical components using a PBS and he clones each component separately. After, he joints the components again to obtain the clones of the polarization state. The process, for $1 \rightarrow M$ cloning machine, is shown in Fig. 2.

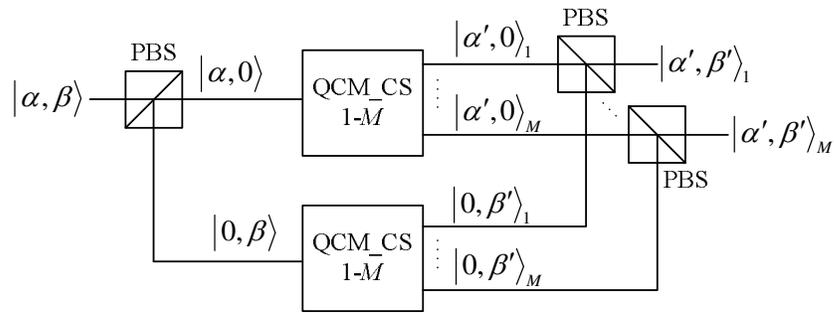

Fig. 2 – Quantum cloning of the polarization of coherent states.

The fidelity of the Gaussian cloning of coherent states does not depend on the mean photon number and it is given by $MN/(MN+M-N)$ [20]. Since the cloning of the polarization requires the cloning of the two components, the fidelity of the polarization cloning is given by $[MN/(MN+M-N)]^2$. In order to the quantum cloning attack to be useful, it is necessary that the fidelity of the clones must be at least larger than the fidelity between one state and the state halfway between two neighbor polarizations. Therefore, the security condition against QCM_CS attack is:

$$\left|\langle\alpha,\beta|R\left(\frac{\theta}{2}\right)|\alpha,\beta\rangle\right|^2 \geq \left|\langle\alpha,\beta|\alpha',\beta'\rangle\right|^2 \Rightarrow \qquad (5)$$

$$\Rightarrow \exp\left[-4\left(\frac{\langle n\rangle}{N}\right)\sin^2\left(\frac{\pi}{8M}\right)\right] \geq \left[\frac{MN}{MN+M-N}\right]^2 \quad 1\leq N \leq M-1 \qquad (6)$$

In (5) $|\alpha',\beta'\rangle$ is the clone of $|\alpha,\beta\rangle$. In (6) we consider that Bob can produce $M$ copies using the state sent by Alice or, he can divide the pulse sent by Alice in $N$ states having mean photon number $\langle n\rangle/N$ and using them to produce $M$ copies. Hence, in order to obtain security, (6) must be satisfied for all $N$ inside the interval $1\leq N \leq M-1$.

At last, the probability of Alice cheating Bob telling him that she choose one of the neighbors of the polarization in fact chosen is given by:

$$p_a = \left\{1-\exp\left[-\mu\langle n\rangle\cos^2(\theta)\right]\right\}\exp\left[-\langle n\rangle\sin^2(\theta)\right] \qquad (7)$$

In (7), $\mu$ is the quantum efficiency of Bob's non-single-photon detector. In the Table 1, considering $P_A$ (equation (4)) uniformly distributed, $\mu=0.75$ and $r_s^1=0.5$, it is shown the values of $\langle n\rangle$, $p_a$, $p_b$ and the security condition against QCM_CS attack for the firsts 12 values of *M*. In particular, the last column of Table 1 receives the value '1' if (6) is satisfied for all values of *N* and '0' if not.

| M | $\langle n \rangle$ | $p_a$ (%) | $p_b$ (%) | Eq. (7) * |
|---|---|---|---|---|
| 2 | 1.183 | 19.832 | 2.928 | 1 |
| 3 | 2.586 | 40.153 | 0.492 | 1 |
| 4 | 4.552 | 48.552 | 0.116 | 1 |
| 5 | 7.081 | 50.438 | 0.034 | 1 |
| 6 | 10.171 | 50.552 | 0.012 | 1 |
| 7 | 13.823 | 50.433 | 0.005 | 1 |
| 8 | 18.036 | 50.333 | 0.002 | 1 |
| 9 | 22.812 | 50.263 | 0.0008 | 1 |
| 10 | 28.150 | 50.213 | 0.0004 | 1 |
| 11 | 34.049 | 50.176 | 0.0002 | 1 |
| 12 | 40.510 | 50.148 | 0.0001 | 0 |

Table 1 – Values of $\langle n \rangle$, $p_a$, $p_b$ and the security condition against QCM_CS attack for the firsts 12 values of *M*. * Value '1'('0') means that equation (6) is (not) satisfied for all *N* in the interval [1,*M*-1]. $P_A$ uniformly distributed, $\mu$=0.75 and $r_s^1$ =0.5.

Observing Table 1, one can see that the proposed protocol is secure against QCM-CS attack only for *M*<12. Moreover, the larger the value of *M* the lower the probability of Bob cheating Alice. On the other hand, the probability of Alice cheating Bob, equation (8), tends to 50%. If $r_s^1$ decreases (increase), neighbors polarizations become more (less) orthogonal, $p_a$ also decreases (increase) but the QCM_CS attack becomes more (less) effective, for example, if $r_s^1$=0.1, only the situations where *M*<4 are secure.

**Conclusions**

We have proposed the use of the polarization of mesoscopic coherent states to realize a bit string commitment protocol. It has been shown that, for *M*<12 and $P_A$ uniformly distributed, $\mu$=0.75 and $r_s^1$ =0.5, the protocol is resistant against brute force and quantum clone machine attacks realized by Bob. On the other hand, Alice can cheat Bob with probability close to 50% for $M \geq 4$. This is a more secure situation than that found in QBC protocols based on single-photon pulses where Alice can cheat Bob with very high probability.

 Acknowledgements

This work was supported by the Brazilian agency FUNCAP.